\documentclass[prd,twocolumn]{revtex4}           

\def\Rhat{\hat R}

\usepackage{amssymb,latexsym}              \catcode`\"=\active \let"=\"

\def\ifundefined#1{\expandafter\ifx\csname#1\endcsname\relax}
\def\bye{\end{document}}   
\long\def\new#1\endnew{{\bf #1}}
\long\def\del#1\enddel{} 

\def\HS#1 {\hspace*{#1pt}} \def\VS#1 {\vspace*{#1pt}}
\def\BC{\begin{center}}    
\def\EC{\end{center}}

\let\0=\over      \let\1=\vec      \def\2{{1\over2}}    \let\3=\ss
\let\4=\underline \let\5=\overline \let\6=\partial      \def\7#1{{#1}\llap{/}}
\def\8#1{{\textstyle{#1}}}         \def\9#1{{\ifmmode{\pmb{#1}}\else\bf#1\fi}}

          \def\({\left(}       \def\){\right)}

\def\eeql#1 {\label{#1}\eeq}      \let\nn=\nonumber  
\def\beq{\begin{equation}}      \def\eeq{\end{equation}}        
\def\bea{\begin{eqnarray}}      \def\eea{\end{eqnarray}}

\let\and=\wedge

\let\bra=\langle        \let\ket=\rangle        \def\<#1\>{\bra #1 \ket}

\def\rel#1 #2{\buildrel #1 \over {#2}}  

          \let\d=\delta   
         \let\th=\theta  
     \let\m=\mu      
\let\n=\nu                       \let\s=\sigma

        \let\L=\Lambda     

   \def\cd{{\cal D}}

\def\cm{{\cal M}}  \def\co{{\cal O}}

\def\IR{{\mathbb R}}

\begin{document}
\title{Inhomogeneity implies Accelerated Expansion}
\preprint{TUW-13-16}
\author{Harald Skarke}
\affiliation{Institut f\"ur Theoretische Physik, Technische Universit\"at Wien,
        Wiedner Hauptstrasse 8-10, A-1040 Vienna, Austria}
\date{\today}

\begin{abstract}
The Einstein equations for an inhomogeneous irrotational dust universe are 
analysed.
A set of mild assumptions, all of which are shared by the standard FLRW type 
scenarios, results in a model that depends only on the
distribution of scalar spatial curvature.
If the shape of this distribution is made to fit the structure of the
present universe, with most of the matter in galaxy clusters and very little 
in the voids that will eventually dominate the volume, then there is a 
period of accelerated expansion after cluster formation, 
even in the absence of a cosmological constant.
\end{abstract}

\maketitle

\section{Introduction}
The celebrated discovery \cite{Riess,Perlm} that the expansion of the 
universe has accelerated in recent times is ususally interpreted as 
indicating the existence of a dark energy or a non-vanishing cosmological
constant, but there have also been many attempts at different explanations
of the observations (see e.g.~\cite{CELU} for an overview).

One particular approach \cite{KMR,Ras,Bu07} relies on the fact that an 
averaging process with respect to a volume based measure need not commute 
with taking a time derivative, thereby resolving the apparent paradox that
there can be an accelerating global expansion even when the local expansion
rate is decelerating everywhere.
This approach relies on the irrotational dust approximation for treating
the geometry of the universe, and on Buchert's formalism \cite{Bu99} for
handling volume averages.
The present work follows this line of reasoning in using the irrotational 
dust approximation.
It deviates, however, by not applying Buchert's equations but instead 
relying on a time independent measure that is proportional to the total
amount of mass in a region.
Along the route a few simplifying assumptions are made, but 
none of them is not shared by the standard 
Friedmann--Lemaitre--Robertson--Walker (FLRW) type scenarios that are
commonly used to infer a non--vanishing value of the cosmological constant.
In fact, our class of models contains the FLRW models as special cases;
what makes the actual universe different is precisely the fact that it
contains both regions whose expansion is halted (galaxies and clusters) and 
regions that continue expanding (voids).

In the following section we present some essential points about the
irrotational dust model that will be required later.
Section 3 contains an exposition of our model and a computation of the leading
and subleading behaviour of the Hubble rate at early and late times.
We find that the Hubble rate behaves asymptotically like that of
an open FLRW universe, but with significant deviations at intermediate times:
for a universe in which most of the matter ends up in bound structures, the 
Hubble rate falls below the FLRW values during the era of structure 
formation by collapse, thus requiring a subsequent period when it approaches 
these values again.
The latter is identified with the present era of accelerated expansion.
In the final section we revisit our assumptions once more;
we find that while it will probably be very hard to use the present model to
make predictions that are as precise as those of the FLRW picture 
(which are wrong, as we argue), the qualitative aspects should be robust.

\section{The irrotational dust universe}
Throughout this note we will model the universe as consisting of irrotational
dust. 
This has the advantage of implying a physically well motivated split of the 
4-manifold ${}^{(4)}\!\cm$ representing spacetime as a cartesian product
\beq {}^{(4)}\!\cm = \IR_+ \times \cm  \eeq
of time and space.
The metric in the synchronous gauge is
\beq ds^2 = -dt^2 + g_{ij}(t,x) dx^i dx^j, \eeql{dustmetric}
where $t$ is the time measured by an observer who is co-moving with the dust. 
The corresponding stress-energy tensor takes the form
${}^{(4)}T_{\m\n}=\rho \d^0_\m\d^0_\n$.
All the information is in the time-dependent spatial 3-metric $g_{ij}(t,x)$, 
so in this note geometric expressions like the Ricci tensor $R_{ij}$ refer 
to the spatial geometry unless explicitly indicated otherwise. 
Indices are lowered and raised with the help
of $g_{ij}$ and its inverse $g^{jk}$, respectively.
Differentiation with respect to $t$ will be indicated by a dot and 
spatial differentiation by subscripts separated by vertical strokes.

We aim for the following. 
Given a domain $\cd \subset \cm$ that is large compared to the
scale of inhomogeneities, we want to compute the Hubble rate as
\beq H_\cd = {\dot V_\cd \0 3 V_\cd} \quad\hbox{ with }
\quad V_\cd(t) = \int_\cd \sqrt{g(x,t)} ~d^3 x.  \eeq
Rather than follow the standard path of using Buchert's equations for 
$\cd$--averaged quantities \cite{Bu99} we prefer to think of $\cd$ as
consisting of many `infinitesimal' (in fact, small in cosmic terms but 
large enough for the irrotational dust approximation) subregions whose 
evolutions we follow independently before adding them up.

With the metric (\ref{dustmetric}) the four--dimensional Einstein tensor 
${}^{(4)}G$ has the components
\bea 
{}^{(4)}\!G^0_0 &=& \2 \th^i_j\th^j_i - \2 \th^2 - \2 R,\\
{}^{(4)}\!G^0_{i} &=& - 2\th^j_{[i|j]},\label{ee0i}\\
{}^{(4)}\!G^i_j &=& 
   \dot\th^i_j + \th \th^i_j - \d^i_j(\dot\th +\2 \th^2 + \2\th^i_j\th^j_i)
   + R^i_j - \2\d^i_jR,
\eea
where the expansion tensor $\th^i_j$ and  the scalar expansion rate $\th$
are defined by
\beq
\th^i_j=\2 g^{ik}\dot g_{kj}, \qquad \th = \th^i_i = {\dot{\sqrt{g}}\0\sqrt{g}} .
\eeql{exptens}
We decompose the expansion tensor and the Ricci tensor into their trace 
and traceless parts,
\beq \th^i_j = {\th\0 3}\d^i_j +\s^i_j, \qquad R^i_j = {R\0 3}\d^i_j +r^i_j,\eeq
hence $\th^i_j\th^j_i = {1\03}\th^2 + 2 \s^2$ with $\s^2 = \2 \s^i_j\s^j_i$.
Then covariant conservation of the stress--energy tensor, 
${}^{(4)}{T_{\m\n ;}}^\n=0$ is equivalent to
\beq \dot\rho + \th\rho =0,  \eeql{rhoev}
and the Einstein equations become
\bea 
{1\0 3}\th^2 - \s^2 + \2 R - \L &=& 8 \pi G_N \rho, \label{einst00}\\
-2\th_{|i} + 3 \s^j_{i|j} &=& 0, \label{einst0i}\\
-2\dot\th - \th^2 - 3\s^2 - \2 R + 3 \L &=& 0, \label{thev}\\
\dot\s^i_j + \th \s^i_j + r^i_j &=& 0, \label{sigoev}
\eea
with the last two equations corresponding to the trace and traceless part of 
${}^{(4)}G^i_j + \L\d^i_j = 0$.
A straightforward computation shows that the time evolution of the
Ricci tensor can be 
described in the notation of Eq.~(\ref{exptens}) as
\beq \dot R_{ij} = \th^k_{i|jk} + \th^k_{j|ik} -\th_{ij|kl}g^{kl}-\th_{|ij}.\eeq
With the vanishing of the r.h.s.~of Eq.~(\ref{ee0i}) 
this implies $g^{ij}\dot R_{ij} = 0$, hence the Ricci scalar $R=g^{ij}R_{ij}$
evolves according to
\beq \dot R + {2\0 3} \th R  = -2\s^i_jr^j_i .\eeql{Roev}
Now we introduce a \emph{local} scale factor 
\beq a(t,x) = a_\mathrm{in}(x) 
     \exp\({1\0 3}\int_{t_{\mathrm in}}^t\th(\tilde t,x)d\tilde t \)
\eeql{lsf}
where the subscript `in' refers to some initial time.
With $\th = 3 \dot a / a$ and rescaled quantities
\beq \hat\rho = a^3 \rho,  ~~~ \hat\s^i_j = a^3 \s^i_j,  ~~~ \hat R = a^2 R,
   ~~~ \hat r^i_j = a^2 r^i_j, \eeql{hatdefs}
the first Einstein equation (\ref{einst00}) becomes a local version 
of the Friedmann equation, 
\beq {\dot a^2 \0a^2} = {\hat\s^2\0 3}a^{-6} + {8\pi\0 3}G_N\hat\rho \,a^{-3}
- {1\0 6}\hat R\,a^{-2} +{\L\0 3},   \label{expev}\eeq
where the evolution of the quantities on the right hand side is determined by
Eqs.~(\ref{rhoev}), (\ref{sigoev}) and (\ref{Roev}) which result in
\beq
\hat{\dot\rho} = 0,\quad
\hat{\dot\s^i_j} = - a \hat r^i_j,\quad
\hat{\dot R} = -2 a^{-3} \hat\s^i_j \hat r^j_i.
\eeql{eveq}
\del
the evolution equations (\ref{rhoev}), (\ref{einst00}), (\ref{sigoev}) and 
(\ref{Roev}) become
\bea 
\hat{\dot\rho} &=& 0,\label{rhev}\\
\dot a^2 &=& {\hat\s^2\0 3}a^{-4}+{8\pi\0 3}G_N\hat\rho a^{-1}-{1\0 6}\hat R
  +{\L\0 3} a^2,   \label{expev}\\
\hat{\dot\s^i_j} &=& - a \hat r^i_j,\label{sigev}\\
\hat{\dot R} &=& -2 a^{-3} \hat\s^i_j \hat r^j_i.\label{Rev}
\eea
\enddel
The first of these equations shows that $\hat \rho = a^3 \rho$ is constant
in time.
We now use our freedom in normalizing $a$, as represented by
the integration constant $a_\mathrm{in}(x)$ in Eq.~(\ref{lsf}), to make
$\hat \rho$ spatially constant as well.
In other words, our conventions are such that the inverse volume $a^{-3}$ 
is proportional to the density $\rho$ with a proportionality factor 
that is constant over spacetime.
The validity of Eq.~(\ref{thev}) is then a consequence of Eqs.~(\ref{expev}) 
and (\ref{eveq}).

\section{A model based on the distribution of curvature}
Up to now we have given an exact description of a completely general 
irrotational dust universe.
The standard FLRW scenario consists in assuming that a good approximation 
of the actual universe is provided by taking $\hat\s=0$, $\hat r = 0$
and $\hat R $ constant in space and time, thus satisfying 
Eqs.~(\ref{eveq}) 
identically by vanishing left and right hand sides. 
Then Eq.~(\ref{expev}) is solved for constant values of $\hat R$ and $\L$,
and the observed acceleration of the expansion is seen as an indication of a
positive $\L$.

Let us examine the impact of these assumptions.
We begin our own analysis by accepting the proposition that 
$\hat r = 0$ provides a reasonable approximation.
Then Eqs.~(\ref{eveq}) imply that $\hat \s$ and $\hat R$
do not evolve in time.
From the structure of Eq.~(\ref{expev}) it is clear that the impact of 
$\hat \s$ is greatest at times when $a$ is small, i.e. in the early universe 
or near collapse. 
As observations seem to indicate negligible influence of shear at early 
times, we
follow the standard procedure of neglecting $\hat \s$ in the following, 
postponing further analysis of this approximation to the discussion section.

What about the assumption of constant $\hat R$?
If this were correct, then we would either have contraction everywhere or 
expansion everywhere. 
Obviously this is in contradiction with the structure of our actual universe
which features both contraction (to galaxies and clusters) and continuing 
expansion (in voids).
Therefore we consider our model universe to be the union of many
infinitesimal regions, each of which has its own $\hat R$ and evolves
according to Eq.~(\ref{expev}) with $\hat \s = 0$.
General statements can then be made by integrating/averaging over these 
regions, according to some probability measure $d\m$ for $\hat R$.

As a consequence of Eqs.~(\ref{exptens}) and (\ref{rhoev}) the expression
$\rho(x,t)\sqrt{g(x,t)}$, 
and therefore the mass content
\beq m_\cd = \int_\cd \rho(x,t)\sqrt{g(x,t)} ~d^3 x  \eeql{mc}
of any domain $\cd\subset\cm$ is time independent.
We assign to any scalar quantity $X(x,t)$ a mass--weighted $\cd$--average as
\beq 
\< X\>_\cd (t) = {1\0 m_\cd} \int_\cd X(x,t)\rho(x,t)\sqrt{g(x,t)} ~d^3 x.
\eeql{xavg}
We emphasize that this is \emph{not} the same as Buchert's average \cite{Bu99};
in contradistinction to the latter, the present averaging prescription
commutes with taking time derivatives, $\< X\>\dot{}_\cd = \<\dot X\>_\cd$.
If $X$ is a function of $t$ and $\hat R(x)$ then 
\beq \< X\>_\cd (t) = \int_{-\infty}^{\infty}X(t,\hat R) ~d\m(\hat R) \eeql{Ravg}
where 
$d\m(\hat R) = [m_\cd^{-1}\int\d(\hat R(x)-\hat R)\rho\sqrt{g}d^3x]d\hat R$
is the fraction of the total 
mass content $m_\cd$ of $\cd$ 
that lies in regions with curvatures in $[\hat R, \hat R + d\hat R]$.
The standard FLRW scenario with a constant rescaled curvature is just
the special case of a discrete distribution with 
$d\m(\hat R)=\d(\hat R-\hat R_\mathrm{const}) d\hat R$.
We can compute the volume of $\cd$ as
\beq V_\cd = \int_\cd \sqrt{g(x,t)} ~d^3 x = m_\cd \< \rho^{-1}\>_\cd 
= {m_\cd\0 \hat \rho}\< a^3\>_\cd ,\eeq
resulting in the Hubble rate
\beq H_\cd={\dot V_\cd\0 3V_\cd}= {\<a^3\>\dot{}_\cd\0 3\<a^3\>_\cd}, \eeql{Hequa}
which can now be analysed in terms of Eq.~(\ref{Ravg}).
In the following we shall omit the subscript $\cd$ from the averages,
with the understanding that we work with a domain $\cd$ that is large enough 
to result in values that are typical for the visible universe.

From now on we restrict ourselves to the case $\L = 0$.
In order to solve Eq.~(\ref{expev}) with $\hat\s^2 = 0$, we change
variables from $a$ to a dimensionless quantity $b$ via
\beq 
a=16\pi G_N\hat\rho~ |\Rhat|^{-1}\,s^2(b) \hbox{ with }
s(b) = \left\{ \begin{array}{ll}\sin(b) & \mbox{if $\hat R > 0$} \\ 
                         \sinh(b) & \mbox{if $\hat R < 0$}  \end{array}
\right. \label{aequa}
\eeq
(the case $\hat R = 0$ is just the limiting case of either of 
the other two).
A brief computation that involves taking the square roots of both sides of 
Eq.~(\ref{expev}) results in
$2|\dot b|s^2(b) = (16\sqrt{6}\,\pi G_N\hat\rho)^{-1}|\Rhat|^{3/2}$
which we can integrate to 
\beq  |b - \2s(2b)| = {|\Rhat|^{3/2} \0 16\sqrt{6}\,\pi G_N\hat\rho}\,t
\eeql{bequa}
(we have set the integration constant to zero in order to match
the convention that time starts at $t=0$).
Our next aim is to obtain asymptotic expressions for the Hubble rate both
for the very early and for the very late universe.
In each case we first eliminate $b$ from Eqs.~(\ref{aequa}) and (\ref{bequa}) 
to obtain an expansion for the local scale factor $a$, 
and then compute $H$ via Eq.~(\ref{Hequa}).
\\
{\bf Early universe.} 
This is the case of $b<<1$.
Inserting the power series for sin(h) 
into Eqs.~(\ref{aequa}) and (\ref{bequa}) 
and eliminating $b$ from the resulting system yields
\bea a(t, \hat R) = \(6\pi G_N\hat\rho ~t^2\)^{1\0 3}
& \( 1- {\Rhat\0 5}\({3\0 2}\,{t\0 16\sqrt{6}\,\pi G_N\hat\rho}\)^{2\0 3}\right.
\nn\\
& \left. +\co\(\Rhat^2(G_N\hat\rho)^{-{4\0 3}}t^{4\0 3}\)\). \eea
We can compare this expression with the quasi-isotropic expansion of 
Lifshitz and Khalatnikov \cite{LK} by using Eq.~(\ref{hatdefs}) to return to
the original quantities $\rho$ and $R$.
This results in 
$6\pi G_N  \rho = t^{-2} + {9\0 40}R+ \ldots$, 
which is identical to what one gets by performing a quasi-isotropic 
expansion for the irrotational dust case \cite{KKS}\footnote{
Redoing the computation that leads from Eq.~(32) to Eq.~(40) of \cite{KKS} 
indicates that the factor of 9 occurring in Eq.~(40) should be replaced by a 
factor of 3; inserting $k=0$, $\protect\tilde{P} = t^{4/3} R$ then gives 
the desired result.}.
Hence our simplifications do not affect the leading orders.

Applying (\ref{Hequa}) leads to 
\beq H(t) = {2\0 3t}\(1-
{\<\Rhat\>\0 5}\({3\0 2}\,{t\016\sqrt{6}\,\pi G_N\hat\rho}\)^{2\0 3}+\cdots\). 
\eeql{Hearly}
This result is consistent with an asymptotic expansion that was performed in
\cite{LS} in the framework of perturbation theory and the Buchert equations.\\
{\bf Late universe.} We distinguish between regions that finally contract,
whose asymptotic volume we model by a constant, and regions that expand forever.
The latter correspond to $\Rhat< 0$ where $s(b)=\sinh(b)\approx e^b/2$ for 
$b>>1$.
By again eliminating $b$ from Eqs.~(\ref{aequa}) and (\ref{bequa}) we find 
the asymptotic behaviour 
\beq a(t, \hat R) = \sqrt{|\Rhat|\0 6}~t
+8\pi G_N\hat\rho|\Rhat|^{-1}
\ln\({|\Rhat|^{3\0 2}t\0 16\sqrt{6}\,\pi G_N\hat\rho}\)+\cdots, \eeq 
for the expanding regions.  
Averaging the third power of this expression and 
adding a constant (which does not contribute to the leading orders
presented here) for the contracted regions results in an expansion for 
$\<a^3\>$ that 
can be inserted into Eq.~(\ref{Hequa}). The result is 
\beq H(t) = {1\0 t} 
\( 1 - {8\sqrt{6}\,\pi G_N\hat\rho \0 \<|\Rhat|^{3\02}\>_-~t}\,
\ln \({\<|\Rhat|^{3\0 2}\>_-t\0 16\sqrt{6}\,\pi G_N\hat\rho}\)+\cdots \),
\eeql{Hlate}
where $\<|\Rhat|^{3\02}\>_-= \int_{-\infty}^0 |\Rhat|^{3\02}~d\m(\hat R)$;
as the subscript indicates, we integrate only over negative values of 
$\Rhat$.
Note that in the last two equations the prefactor of $t$ under the logarithm 
is inessential because it can be modified by lower order terms;
it was chosen to conform to Eq.~(\ref{bequa}).

Let us denote by $\Rhat_c$ a scale that is characteristic for the 
distribution of $\Rhat$ so that 
\beq \<\Rhat\>=\co(\Rhat_c)\quad \hbox{ and }\quad 
\<|\Rhat|^{3\02}\>_-=\co(|\Rhat_c|^{3\02}). \eeq
Then Eqs.~(\ref{Hearly}) and (\ref{Hlate}) imply that $H(t)$
is a function that behaves
asymptotically like ${2\03}\,t^{-1}$ at early times and like
$t^{-1}$ at late times, with a transition era around 
a characteristic time scale $t_c= 16\sqrt{6}\,\pi G_N\hat\rho |\Rhat_c|^{-3/2}$.
At leading order this is just the behaviour of an open FLRW universe.
Indeed, such a universe corresponds to the discrete distribution with
$d\m(\Rhat) = \d(\Rhat-\Rhat_c)d\Rhat$ for $\Rhat_c < 0$.
Then $\<\Rhat\> = \Rhat_c < 0$, the subleading term in Eq.~(\ref{Hearly}) is 
positive, the graph of $H(t)$ is between the graphs of the functions  
${2\03}\,t^{-1}$ and $t^{-1}$ everywhere, and $H(t)$ is monotonically 
decreasing even though $tH(t)$ increases from $2/3$ to $1$.

What about a universe that looks more like ours, i.e.~one that features 
collapse?
According to \cite{FHP} the greatest part of the baryon content of the universe
resides in clusters of galaxies, which correspond to regions of positive $\Rhat$
in our model.
Hence we should think of the distribution for $\Rhat$ as heavily skewed towards
positive values, for example a Gaussian distribution with mean 
$\<\Rhat\> = \Rhat_c>0$
and standard deviation of a similar or smaller magnitude.
Then Eq.~(\ref{Hearly}) shows that the graph of $H(t)$ goes \emph{below}
that of ${2\0 3}t^{-1}$.
Indeed, a region characterized by some positive value of $\Rhat$
collapses, according to Eq.~(\ref{aequa}), during the interval 
$b\in(\pi/2,\pi)$, 
i.e.~$(16\sqrt{6}\,\pi G_N\hat\rho)^{-1}|\Rhat|^{3/2}\,t\in(\pi/2,\pi)$.
If the distribution is dominated by the region around some $\Rhat_c>0$ then we
can even get $H(t)<0$ for $t/t_c \approx 3\pi / 4$.
But a negative Hubble rate at a finite time and a positive Hubble rate
at very late times would imply 
an intermediate period of a rising Hubble rate $\dot H > 0$, which is 
more than we need to achieve a negative deceleration parameter 
$q = -1-\dot H/H^2$, i.e.~accelerating expansion.
The following summary of the present paragraph is also the main statement of
this paper.

\emph{In a universe in which a large fraction of the matter content ends up 
in bound clusters and very little matter goes into voids that eventually come 
to dominate the 
total volume, one would expect to have a period of accelerating 
expansion after most of the collapse has taken place, even in the absence 
of a cosmological constant.}

\section{Discussion}

Except for the supposition that the universe contains regions of negative
spatial curvature as well as regions of positive spatial curvature, 
we have not made a single assumption that is not shared by the standard FLRW 
scenario.
This resulted in a model which, when adjusted to the way matter is distributed,
\emph{predicts} that the change in the Hubble rate should be above that for a 
FLRW universe with $\L=0$ in the era after structure formation.
This model rests upon two pillars, the irrotational dust approximation for
the matter content and the neglect of shear.
Let us consider once more the impact of these assumptions.

{\bf The irrotational dust approximation.}
This approximation breaks down in the early universe where the stress--energy
tensor has a non--negligible pressure term, and in collapsing regions
where vorticity starts to play a role and eventually holds the collapse.
While the breakdown in the early universe makes it harder to connect our model 
to data of the cosmic microwave background, it seems extremely unlikely that 
it could affect results that concern the period after structure formation.
The occurrence of vorticity certainly leads to a strong modification of
the way collapse happens in detail, but this need not worry us since 
we did not use any of these details in our treatment of collapsing regions
in the previous section.

On the other hand, the irrotational dust approximation has the great 
advantage of supplying us with a natural, physically well motivated split of
spacetime into space and time, thereby giving a precise meaning to a statement
such as `at time $t$ the volume of domain $\cd$ is $V_\cd(t)$.'

{\bf Neglect of shear.}
This is certainly the greatest weakness of the present model.
Indeed by Eq.~(\ref{einst0i}) vanishing shear would imply a spatially constant
expansion rate.
Therefore our model cannot represent an exact solution to the Einstein 
equations, but that does not prevent it from providing a better approximation
to our universe than any FLRW scenario.

Let us briefly examine what happens when we relax this assumption.
Once we have shear, we must also allow for time dependence of $\hat R$.
As a consequence of Eqs.~(\ref{eveq}) we get
\bea \hat\s^2(t) = 
\hat\s^2(t_{\mathrm{in}})-\int_{t_{\mathrm{in}}}^ta(\tilde t)\,\hat\s^i_j(\tilde t)
\hat r^j_i(\tilde t)\,d\tilde t,\\
\hat R(t) = 
\hat R(t_{\mathrm{in}})-2\int_{t_{\mathrm{in}}}^ta^{-3}(\tilde t)\,\hat\s^i_j(\tilde t)
\hat r^j_i(\tilde t)\,d\tilde t, \eea
where we have suppressed the space dependences and $t_\mathrm{in}$ denotes 
some initial time.
Then we must modify the right hand side of Eq.~(\ref{expev}) by replacing
the formerly constant $\hat \s^2$ and $\hat R$ by their values
at time $t_\mathrm{in}$ and adding the term
\beq {a^{-2}\0 3}\int_{t_{\mathrm{in}}}^t\({1\0 a^3(\tilde t)}
-{a(\tilde t)\0 a^4(t)}\)
\hat\s^i_j(\tilde t) \hat r^j_i(\tilde t)\,d\tilde t   \eeq
which we now want to analyse.
Unless we 
consider collapse, $1/a^3(\tilde t)$ will be
larger than $a(\tilde t)/ a^4(t)$, so we can neglect the latter.
Assuming that the Ricci tensor is of the same order of magnitude as its trace,
so that the components of $\hat r^j_i$ will be limited by a value near our 
typical scale $\hat R_c$, 
the integral will not be much larger than
$\hat R_c \int_{t_{\mathrm{in}}}^t a^{-3}(\tilde t) |\hat\s(\tilde t)| d\tilde t
= \hat R_c \int_{t_{\mathrm{in}}}^t |\s(\tilde t)| d\tilde t$.
For moderate shear this means that we should view this term
as a slowly varying addition to the $\hat R_{\mathrm{in}}$--term.

Our rough estimates indicate that the only modification required to
make our model exact would be to allow for a slow evolution of $\hat R$.
While this invalidates
the assumption that a single distribution $d\m(\Rhat)$ can give
quantitatively precise results for all time scales,
the function $H(t)$ will nevertheless osculate ${2\0 3}t^{-1}$ at early 
times and $t^{-1}$ at late times;
however, the subleading terms of Eqs.~(\ref{Hearly}) and (\ref{Hlate})
may depend on expectation values $\<\Rhat\>_\mathrm{early}$ and 
$\<|\Rhat|^{3\02}\>_{-~\mathrm{late}}$, with the sole difference that the two 
distributions need not be exactly the same.
This certainly makes it harder to produce precise quantitative predictions
but should not affect the qualitative result that 
inhomogeneity affects the shape of the function $H(t)$.
After all, most of the matter in the universe did coalesce in clusters,
so the Hubble rate must have gone below that for an open FLRW universe,
hence there must be a period during which it returns.


\bye